\documentclass[11pt,aps,prc,superscriptaddress,showpacs,floatfix]{revtex4-1}
\usepackage[dvipdfmx]{graphicx} 
\usepackage{epsfig}
\usepackage{bm}
\usepackage{amssymb}
\usepackage{wrapfig}
\usepackage{ulem}
\usepackage{multirow}
\usepackage{color}

\newcommand{\psla}{\ooalign{\hfil/\hfil\crcr{p}}}

\newcommand{\epsisla}{\ooalign{\hfil/\hfil\crcr{$\epsilon$}}}

\newcommand{\Ubar}{\mbox{$\overline{U}$}}

\newcommand{\tldu}{\mbox{${\tilde u}$}}

\newcommand{\vp}{\mbox{$\bm{p}$}}
\newcommand{\vq}{\mbox{$\bm{q}$}}
\newcommand{\vbr}{\mbox{$\bm{r}$}}
\newcommand{\vk}{\mbox{$\bm{k}$}}

\newcommand{\vA}{\mbox{$\bm{A}$}}

\newcommand{\vQ}{\mbox{$\bm{Q}$}}

\newcommand{\vepsi}{\mbox{\boldmath $\epsilon$}}

\newcommand{\br}{\bf{r}}
\newcommand{\bp}{\bf{p}}
\newcommand{\bq}{\bf{q}}

\newcommand{\bk}{\bf{k}}

\begin{document}

\title{Compton Scattering of Hermite Gaussian Wave $\gamma$-Ray}

\author{Tomoyuki~Maruyama}
\affiliation{College of Bioresource Sciences, Nihon University,
Fujisawa 252-8510, Japan }
\affiliation{National Astronomical Observatory of Japan, 2-21-1 Osawa, Mitaka, Tokyo 181-8588, Japan}
\author{Takehito~Hayakawa}
\affiliation{National Institute for Quantum and Radiological Science and Technology, Tokai, Ibaraki 319,-1106, Japan}
\affiliation{National Astronomical Observatory of Japan, 2-21-1 Osawa, Mitaka, Tokyo 181-8588, Japan}
\author{Toshitaka~Kajino}
\affiliation{Beihang University, School of Physics and 
Nuclear Energy Engineering,\\
Int. Center for Big-Bang Cosmology and Element Genesis, Beijing 100191,
China} 
\affiliation{National Astronomical Observatory of Japan, 2-21-1 Osawa, Mitaka, Tokyo 181-8588, Japan}
\affiliation{The University of Tokyo, Bunkyo-ku, Tokyo 113-0033, Japan}
%

%\pacs{12.60.Fz, 32.80.Cy}

%\pacs{12.60.Fz, 32.80.Cy}

\begin{abstract}

We calculate the differential cross sections for Compton scattering of photons described by Hermite Gaussian (HG) wave function in the framework of relativistic quantum mechanics.
The HG wave $\gamma$-rays propagating along the $z$-direction have quantum numbers of nodes of $n_x$ and $n_y$ in the $x$- and $y$-directions, respectively.
The calculated differential cross section is symmetric with respect to both the $zx$- and $zy$-planes.
The nodes whose number is identical with $n_x$ and $n_y$ appear in the energy spectrum measured in $zx$- and $zy$-planes, respectively.
These results indicate that it is possible to identify the HG wave photon and its quantum numbers $n_x$ and $n_y$ by measuring Compton scattering.
The present proposed method can be also applied to $\gamma$-ray astronomy.

\end{abstract}

\maketitle

%%Introduction
In 1992, Allen {\it et al.} \cite{Allen92} presented that a single photon with a topological wave function such as Laguerre Gaussian (LG)  function can have non-zero orbital angular momentum projection along the photon propagation axis.
%The total angular momentum is larger than or equal to the projection of the anuglar momentum.
The LG photon with a helical wave front, i.~e.~so-called ``photon vortex'' has been studied for fundamental science and applications \cite{Molina-Terriza06}.
Furthermore, the concept of the vortex has been extended to various beams such as electrons \cite{Uchida10, Bliokh17} and neutrons \cite{Clark15,Afanasev17}.
One of features of the topological photons with large angular momentum is the fact that the interaction with materials such as atoms \cite{Afanasev13, Schmiegelow16}, nuclei \cite{Taira17, Afanasev17}, and elemental particles \cite{Ivanov11} is different from that of the plane wave photons.
This is because possible excitations or reaction channels are limited by the conservation law of angular momentum \cite{Afanasev13}.
For example, various excitation modes of an atom were demonstrated using vortex laser and an ion trap \cite{Schmiegelow16}.
It was theoretically pointed out that the dominant excitation mode of giant dipole resonance on even-even nuclei in photonuclear reactions was forbidden for $\gamma$-ray vortices \cite{Taira17}.
The differential cross sections of photodisintegration reactions on deuterons with photon vortices weere calculated \cite{Afanasev17}.
Furthermore, high energy photon vortices also show new interactions in high energy physics \cite{Ivanov11}.
In addition, probability of generation of photon vortices in the universe has been discussed \cite{Harwit03,Elias08,Berkhout08,TTMA11}.

The LG photons can be transformed from Hermite Gaussian (HG) wave photons.
The HG function can be mathematically represented by a linear combination of LG functions with different orbital angular momenta and radius quantum numbers \cite{Abramochkin91}.
Therefore, an HG wave photon could have non-zero orbital angular momentum although it does not have a helical wave front.
This leads to the fact that reaction channels on nuclei (or elemental particles) with the HG photons with large angular momentum are different from those with the plane wave photons.
Thus, the HG photon is as useful as the LG photon to study the non-standard interactions with materials.

The HG photons could be generated using high-harmonic radiation from planar undulators with high energy electrons \cite{Sasaki08}.
Another candidate is the generation by inverse Compton scattering with HG laser; high power HG mode laser has been developed \cite{Chu12,Kong12}.
However, there is a critical problem that optical devices to measure topological light at visible wavelengths cannot work in the MeV energy region.
One of the candidates for measuring the HG $\gamma$-rays is to use Compton scattering.
For photon vortices, Compton scattering was calculated in non-relativistic \cite{Stock15} and relativistic quantum mechanics \cite{Jentschura11a, Jentschura11b, Maruyama17}. 
The differential cross section of Compton scattering with the LG photon is axial symmetric along the photon propagation direction, and therefore the azimuthal angle dependence is unity.
Thus, we have proposed the coincidence measurement of the scattered photon and electron from each Compton scattering to identify the LG photon \cite{Maruyama17}.
In contrast, the cross section with linearly polarized $\gamma$-rays is symmetric with respect to the polarization plane,
and they depend on the azimuthal angle between the linear polarization plane and the Compton scattered plane.
In fact, the linear polarization spectrometer based upon Compton scattering has been used for the study of nuclear physics \cite{Jones95} and for observation of astronomical $\gamma$-rays  \cite{Coburn03,Kalemci07}.

Here, we would like to point out that the HG wave function is symmetric with respect to two planes of $zx$- and $zy$-planes as discussed later.
This fact leads to the possibility that the cross section of Compton scattering of the HG photon is also symmetric with respect to these two planes.
If so, it is possible to identify the HG photon by measuring only its azimuthal angle dependence.
In the present Letter, we calculate the differential cross section of Compton scattering of an HG photon on a rest electron in the framework of relativistic quantum mechanics.
We also discuss possible measurements of HG $\gamma$-rays in the laboratory and the universe.

%%%%%%%% Calcualtion

\begin{figure}[htb]
\begin{center}
\vspace{-1em}
{\includegraphics[scale=0.8]{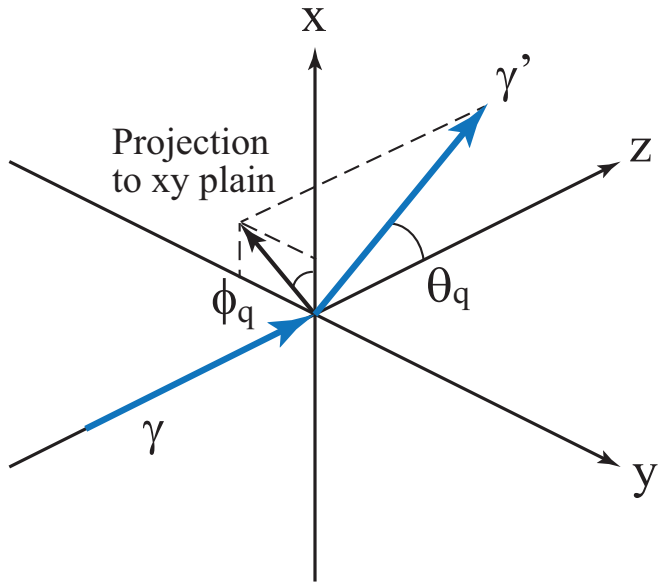}}
\caption{\small
Coordinate system used in the calculation. $\gamma$ and $\gamma$' denote the initial and final photons, respectively.
}
\label{Fig1}
\end{center}
%\end{figure}
\end{figure}

%%%%%%%%% Hermite-Gaussian wave photon
When an initial HG wave photon propagates along the $z$-axis and its wave function has nodes of $n_x$ and $n_y$ in the $x$-direction and $y$-direction, respectively  (see Fig.~\ref{Fig1}), the HG wave function is written as
\begin{eqnarray}
u (\vbr)  &=&  \sqrt{\frac{2}{R_z} }
 \frac{1}{w (z)}
 f_{n_x} \left( \frac{\sqrt{2} x}{w (z)} \right)
 f_{n_y} \left( \frac{\sqrt{2} y}{w (z)} \right)
 \exp \left[ ikz + \frac{ik r^2}{2R(z)}  - i (n_x + n_y  +1 ) \theta_z \right] ,
\label{Eq:HG1}
\end{eqnarray}
with
\begin{eqnarray}
&& f_n (x) = (2^n \sqrt{\pi} n! )^{-1/2} H_n (x) e^{-x^2/2} ,
\nonumber \\ 
&& w (z) = w_0 \sqrt{1 + \frac{z^2}{z_R^2}}, \quad
R(z) = (z^2 + z_R^2)/z,  \quad  
\theta_z =  \tan^{-1} \left(\frac{z}{z_R}\right),
\quad z_R = k w_0^2 /2.
\label{Eq:HG2}
\end{eqnarray}
where $k$ is the energy of the initial photon, $H_n$ is the $n$-th Hermite polynomial, $R_z$ is the size of the system along the $z$-axis, and $w_0$ is the waist radius at $z=0$.

%%% Calculation
We consider Compton scattering of an HG wave $\gamma$-ray on an electron at rest, where
we do not observe the electron spins and the polarization of the final photon. 
We choose the Lorentz Gauge and  $A_0 =0$ for the photon field.
We set the final photon wave function to be the plane wave with momentum of $q \equiv (|\vq|, \vq)$.
%%% electron and photon fields
The electron and photon fields are written as
\begin{eqnarray}
%%%&& \psi(x) = \frac{1}{\sqrt{\Omega}}U(\vp, s) e^{i\bp \br - i E_p t} ,
&& \psi(x) = \frac{1}{\sqrt{\Omega}}U(\vp, s) e^{i{\bp r}  - i E_p t} ,
~ %\nonumber \\ && 
\vA_i(\vbr) =  \frac{\vepsi_i (h_i)}{\sqrt{2 k}} 
 u (\vbr) e^{ - i k t} ,
~
\vA_f (\vbr) =  \frac{\vepsi_f (h_f)}{\sqrt{2 |\vq| \Omega}}
%%% e^{i \bq \br - i |\bq| t} ,
 e^{i {\bq} {\br} - i |{\bq}| t} ,
%%s
\end{eqnarray}
where 
$\Omega$ is the volume of the system, $U(\vp,s)$ is the Dirac spinor 
of an electron with the momentum $p = (Ep, \vp$) and the spin $s$,  and
$\vepsi_{i(f)} (h_{i(f)})$ is the polarization vector with the helicity 
$h_{i(f)}$.
In addition, we write  the initial and final momenta of the electron as 
$p_i = (m, {\bm 0})$ and $p_f = (E_f, \vp_f)$, respectively.
The scattering amplitude \cite{Bjorken} is rewritten as 
\begin{eqnarray}
S_{if} & =& \frac{e^2}{2 \sqrt{k |\vq|  \Omega}}
\Ubar (\vp_f,s_f) \left[ \epsisla_f S_F (p_f + q) \epsisla_i
 +  \epsisla_i  S_F (p_i - q)  \epsisla_f
\right] U(\vp_i,s_i)
\nonumber \\ &&  \qquad\qquad\qquad \times
 \tldu (\vp_f + \vq)  (2 \pi) \delta (E_f + |\vq| -m - k) ,
\end{eqnarray}
with $\epsilon_{i,f} = (0, \vepsi_{i,f})$, where $S_F$ and $\tldu(\vp)$ are defined as
\begin{eqnarray}
&& S_F(p) = \frac{\psla +m}{p^2 +m^2 + i \delta} , \quad
\tldu (\vk) = \int d \vbr e^{-i \bk \cdot \br} u(\vbr).
\label{eq5}
\end{eqnarray}
Then, the cross-section is given  by
\begin{eqnarray}
&& d \sigma =
\frac{e^4}{4 k m} 
 W_{if} \left| \tldu (\vp_f + \vq) \right|^2
(2 \pi) \delta (E_f + |\vq| - m - k )
\frac{d \vq}{(2 \pi)^3 |\vq|} \frac{d \vp_f}{(2 \pi)^3 E_f}
\label{dCrs}
\end{eqnarray}
with
\begin{eqnarray}
W_{if}&=& \frac{mE_f}{2}  \sum_{s_i, s_f, h_f} \left| \Ubar (\vp_f,s_f) 
\left[ \epsisla_f (h_k)  S_F (p_f + q) \epsisla_i + 
\epsisla_i S_F (p_i - q)  \epsisla_f(h_k) \right] U({\bm 0}, s_i) \right|^2 .
\end{eqnarray} 

%%%%% Fourier transformation of u
The Fourier transformation of $u(\vbr)$ in Eq.~(\ref{eq5}) becomes
\begin{eqnarray} 
\tldu(\vQ)
 &=&
\frac{2 \sqrt{2} \pi^2 w_0}{\sqrt{R_z}}
f_{n_x} \left[ \frac{w_0 Q_x }{\sqrt{2} } \right] 
f_{n_y} \left[ \frac{w_0 Q_y }{\sqrt{2} } \right] 
\delta \left( Q_z - k + \frac{Q_T^2}{2k} \right) ,
%%%
\label{PhAmp}
\end{eqnarray}
%%/
with $Q_T^2 = Q_x^2 + Q_y^2$.
The delta-function in Eq.~(\ref{PhAmp}) indicates the condition of
 $Q_z = k - Q_T^2/2k$,
which is derived from the on-mass-shell condition, $k = |\vQ|$, 
with the para-axial approximation of $k \approx Q_z \gg Q_T$.
We improve this term by satisfying the exact kinematical
condition of $Q_z = \sqrt{k^2 - Q_T^2}$.
In addition, we  use  the polarization vector of the initial photon 
to satisfy the exact relations: $\vepsi_i \cdot (\vp_f + \vq ) = 0$. 
As the result, the  amplitude $W_{if}$ becomes
\begin{eqnarray}
&& W_{if}  =
\frac{4 |\vq|}{k} + \frac{4 k}{|\vq|} - \frac{4}{k^2} 
\left[|\vp_f|^2 - \frac{(\vp_f \cdot \vq)^2}{|\vq|^2} \right] .
\end{eqnarray}
%Note that the abve $W_{if}$ is more simple than that in
%Ref.~\cite{LGCmp} owing to the two improvement for the on-mass-shell
%condition and the polarization vector. 
%However, the two improvements do not change numerical results 
%n our work where the para-axial approximation is available.
%%% The final cross sections
We calculate the photon cross-section integrated over the final electron momentum.
We finally obtain the cross section
\begin{eqnarray}
\frac{d^3 \sigma}{d \Omega d E_q} &=& 
\frac{\alpha^2 w_0^2 E_q}{2 m k }\int  \frac{d \vp_f}{E_f} 
\delta  (E_f + E_q - m -k) 
\delta \left( Q_z -\sqrt{k^2 - (Q_x^2 + Q_y^2)} \right) W_{if}
\nonumber \\ && \qquad\qquad\qquad \times
 \left[f_{n_x} \left(\frac{w_0 Q_x }{\sqrt{2} } \right) 
f_{n_y} \left( \frac{w_0 Q_y }{\sqrt{2} } \right) \right]^2 ,
\label{dCrs2}
\end{eqnarray}
where $E_q = |\vq|$ and $\vQ = \vp_f + \vq$.

%%% Parameters w0, k
To calculate quantitatively the cross sections, we assume that HG wave photons with an energy $k$ of 500 keV propagate along the $z$-axis.
The HG wave function has a waist and spread beyond the waist at $z$ = 0.
The opening angle of the HG wave is defined by the photon energy $k$ and the waist radius $w_0$ [see Eqs.~(\ref{Eq:HG1}) and (\ref{Eq:HG2})].
Optical devices to focus X/$\gamma$-rays have been developed and thereby it is possible to focus 20 keV X-rays on a small area with a diameter of 7 nm at present \cite{Miura09};
the ratio of this diameter to its wave length is approximately 100.
Applying this ratio to 500 keV (2.48 pm), the diameter of the focused area is approximately 250 pm.
Thus, in the present calculation, we take the waist radius $w_0$ as 25 pm, 75 pm and 250 pm to study the $w_0$ dependence.
Furthermore, it is expected that the HG photons are directly generated by fundamental processes such as high-order harmonic radiations from planar undulators with electron beams at energies of GeV \cite{Sasaki08} and the inverse Compton scattering on relativistic electrons with HG laser. 
Because a single electron radiates a single photon in these processes, it is expected that individual electrons directly generate the HG wave photons under specific conditions in which their waist radius probably correlates with their wave length.
Thus, the present assumed waist radius are not unrealistic values.  

\begin{figure}[htb]
\begin{center}
%\vspace{-1.5em}
{\includegraphics[scale=0.7, angle=-90]{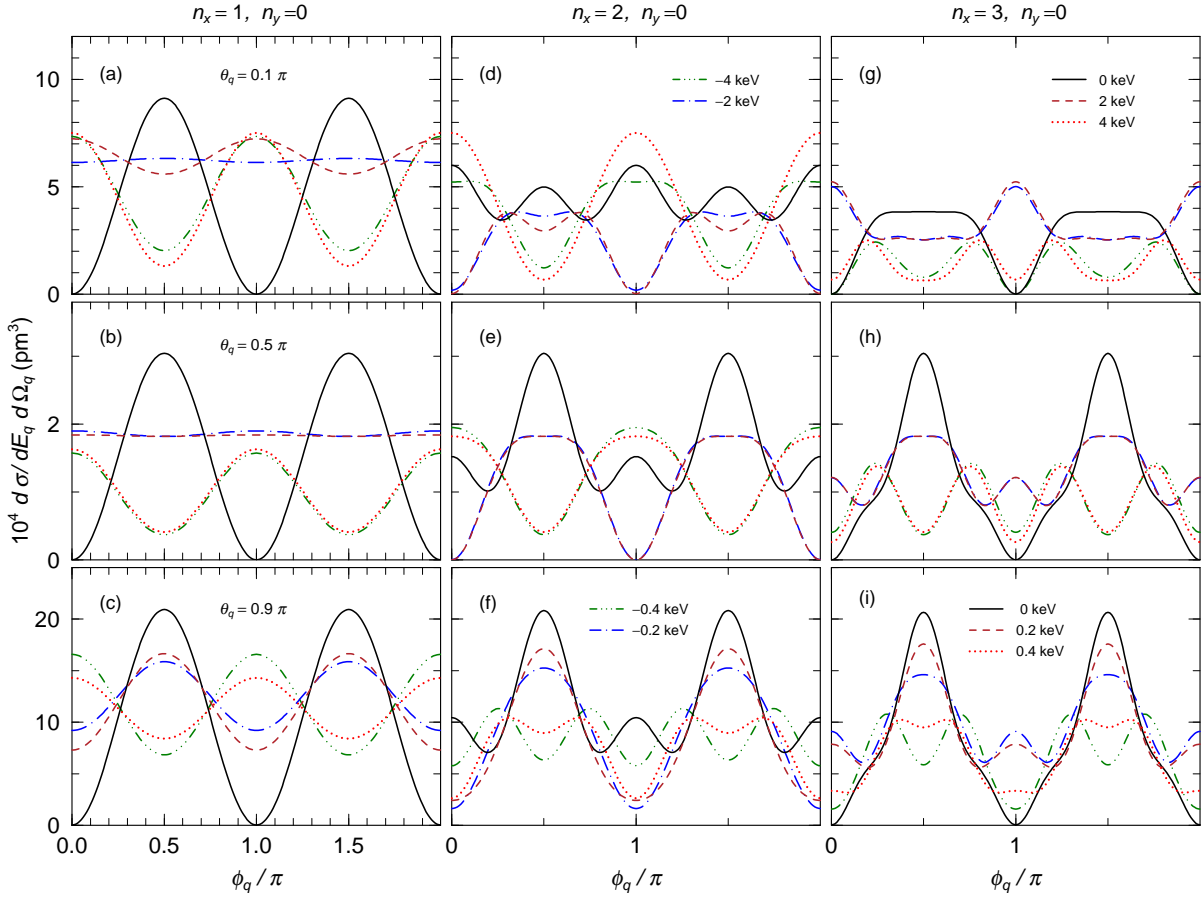}}
\caption{\small
The differential cross sections of Compton scattering with HG wave function photons of the waist radius $w_0$ = 25 pm
for $n_x$ = 1 and $n_y$ = 0 (a,~b,~c),  $n_x$ = 2 and $n_y$ = 0 (d,~e,~f), and  $n_x$ = 3 and $n_y$ = 0 (g,~h,~i).
The scattered polar angles are $\theta_q$ = 0.1 $\pi$ (a,~d,~g),  $\theta_q$ = 0.5 $\pi$ (b,~e,~h), and  $\theta_q$ = 0.9 $\pi$ (c,~f,~i).
The solid (black), dashed (dark red), dot (red), dot-dashed (blue), two-dot-dashed (green) lines indicate the cross sections for $\delta$E = 0 keV, 2 keV, 4 keV, -2 keV, -4 keV, respectively, for $\theta_q$ = 0.1 $\pi$ and 0.5 $\pi$.
The solid (black), dashed (dark red), dot (red), dot-dashed (blue), two-dot-dashed (green) lines indicate the cross sections for $\delta$E = 0 keV, 0.2 keV, 0.4 keV, -0.2 keV, -0.4 keV, respectively, for $\theta_q$ = 0.9 $\pi$.
}
\label{Fig2}
\end{center}
\end{figure}

%%%% Result for azimuthal dependence 
Figure \ref{Fig2} shows the differential cross sections as functions of  the polar angle $\theta_q$ for various azimuthal angles $\phi_q/\pi$ and various node numbers of $n_x$ and $n_y$.
In the case of standard Compton scattering of a plane wave photon, the scattered photon energy, $E_0$, is uniquely determined when the scattered angle is fixed because of the conservation law of momentum. 
In contrast, the scattered photon energy for the incident HG wave photon may shift from that for the standard Compton scattering.
Thus, we present the differential cross sections for various energy difference between the energies for the HG photon and the plane wave photon ($\Delta$E = $E_q$ - $E_0$).
The solid lines in Fig.~\ref{Fig2} show the cross sections for $\Delta$E = 0.
In the case of $n_x$ = 1 and $n_y$ = 0, the $\phi_q$ dependence of the cross sections for $\Delta$E = 0 at the three polar angles of $\theta_q$ = 0.1$\pi$, 0.5$\pi$, and 0.9$\pi$ is almost identical.
This $\phi_q$ dependence is similar to that for Compton scattering with linearly polarized $\gamma$-rays.
In contrast, in the cases of  $n_x$ $\ge$ 2, the cross sections for $\Delta$E = 0 are different from that in  $n_x$ = 1.
These results indicate that it, in principle, is possible to distinguish the HG wave photon of $n_x$ $\ge$ 2  from the linearly polarized photons and to identify $n_x$ and $n_y$.
%

\begin{figure}[htb]
\begin{center}
%\vspace{-1.5em}
{\includegraphics[scale=0.7]{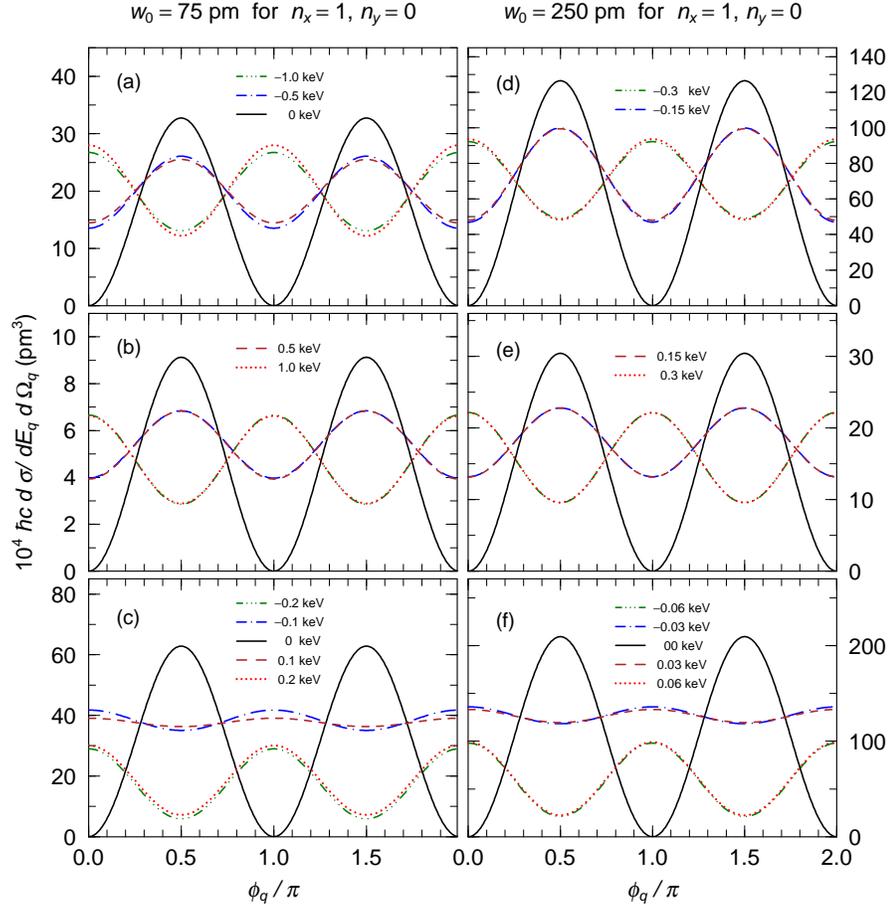}}
\caption{\small
The differential cross sections of Compton scattering with HG wave function photons for $n_x$ = 1 and $n_y$ = 0 in the cases of the waist radius $w_0$ = 75 pm and 250 pm.
The scattered polar angles are $\theta_q$ = 0.1 $\pi$ (a,~d),  $\theta_q$ = 0.5 $\pi$ (b,~e), and  $\theta_q$ = 0.9 $\pi$ (c,~f).
The solid (black), dashed (dark red), dot (red), dot-dashed (blue), two-dot-dashed (green) lines indicate the cross sections for $\delta$E = 0 keV, 0.5 keV, 1keV, -0.5 keV, -1 keV, respectively, for $\theta_q$ = 0.1 $\pi$ and 0.5 $\pi$.
The solid (black), dashed (dark red), dot (red), dot-dashed (blue), two-dot-dashed (green) lines indicate the cross sections for $\delta$E = 0 keV, 0.1 keV, 0.2 keV, -0.1 keV, -0.2 keV, respectively, for $\theta_q$ = 0.9 $\pi$.
}
\label{Fig3}
\end{center}
\end{figure}

%%% Energy shifts
As stated above, the energy of the scattered photon may be shifted from that for the standard Compton scattering.
As shown in Fig.~\ref{Fig2}, as $\Delta$E increases, the differential cross sections drastically change.
Thus, it is required to determine $\Delta$E in order to identify the HG photon.
The range of the energy shift depends on the polar angle $\theta_q$ of the scattered photon; as $\theta_q$ increases, the shift decreases.
This indicates that by measuring the energy of the scattered photon at forward angles when the incident photon energy is known,  the HG photon can be relatively easily identified.
%%% Experiment in laboratory
Even if the incident energy is unknown, by measuring simultaneously the deposited energy in the primary active target where Compton scattering occurs and the energy of the scattered photon in the secondary detector, it is possible to know both the energy of the incident photon and the energy shift $\Delta$E. 
This experimental technique is being commonly used in nuclear physics and $\gamma$-ray astronomy.
To investigate the waist radius dependence, we present the cross sections for $w_0$ = 75 pm and 250 pm in Fig.~\ref{Fig3}. 
We find that the results in these two cases are similar to that with $w_0$ = 25 pm although the absolute value of the energy shift decreases as $w_0$ increases.
The ratio of the energy shift to the expected energy is in the range of $\Delta$E/E = 0.001$\--$0.01 except for the case of $w_0$ = 250 pm and $\theta_q$ = 0.9$\pi$. 
Because the typical energy resolutions of semi-conductor detectors are approximately 0.001$\--$0.002 in full width at half maximum, the energy shifts in most of currently assumed conditions could be measured. Furthermore, even if $w_0$ $>$ 250 pm, it is possible to measure the cross sections at forward angles.

\begin{figure}[htb]
\begin{center}
%\vspace{-1.5em}
{\includegraphics[scale=0.7, angle=-90]{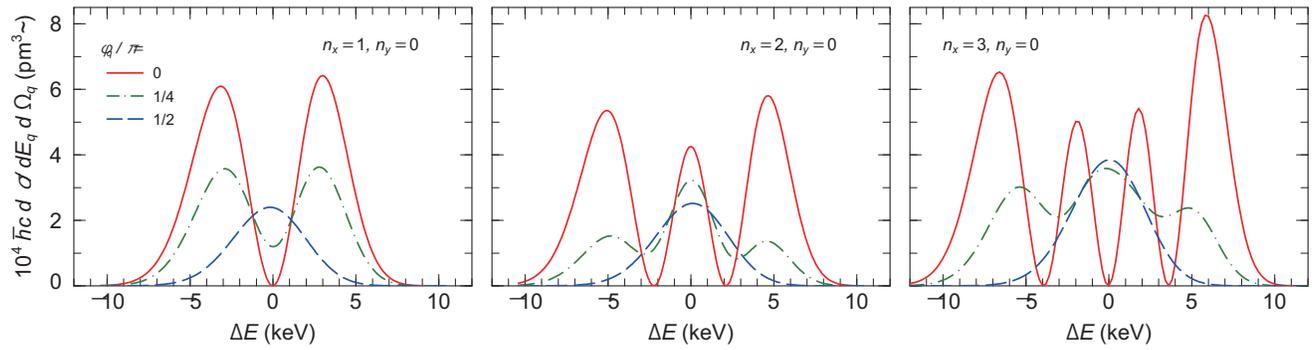}}
\caption{\small
The energy spectra of the scattered photons with HG wave of the waist radius $w_0$ = 75 pm in the case of $\theta_q$ = 0.1$\pi$. The solid  (red) , dot-dashed (green), and dashed (blue) lines indicate that in the case of $\phi_q$ = 0, 1/4 $\pi$, and 1/2 $\pi$, respectively, where $\phi_q$ is the azimuthal angle from the $zx$-plane.
}
\label{Fig4}
\end{center}
\end{figure}

%%% Nodes in energy spectra
Figure \ref{Fig4} shows the energy spectra of the photons scattered at $\theta_q$ = 0.1$\pi$.
One can see the nodes in the energy spectra at $\phi_q$ = 0 ($zx$-plane).
It should be noted that the number of the observed nodes is identical with $n_x$.
The energy spectra change as the azimuthal angle increases, and thereby the nodes disappear at $\phi_q$ = 1/2$\pi$ ($zy$-plane) because of $n_y$ = 0.
This indicates that the measurement of the energy spectra as functions of the azimuthal angle provides the quantum numbers $n_x$ and $n_y$ of the initial HG photon.

%%% Gamma-ray astronomy
Gamma-ray bursts (GRBs) are one of the most energetic explosive phenomena in the universe and the mechanism has been an unresolved problem.
The linear polarization of photons from GRBs was measured using polarimeters based upon Compton scattering in satellites.
The observed polarization is as high as 80\% measured by the RHESSI satellite \cite{Coburn03} and 98\% by SPI/INTEGRAL \cite{Kalemci07}.
%%%The telescopes in these satellites consist of multi-segmented $\gamma$-ray detectors.
The highly polarized $\gamma$-rays are considered to be generated by synchrotron radiations from relativistic electrons under strong magnetic fields in stars or by inverse Compton scattering.
This gives a speculation that HG wave $\gamma$-rays may be generated by high-order harmonic generations under strong magnetic fields. 
If so, as discussed previously a multi-segmented $\gamma$-ray detector system could identify HG $\gamma$-rays from GRBs.
Some polarimeters consist of two layers of segmented-detector arrays; the energy deposited by Compton scattering is measured by one of the first layer detectors and simultaneously the scattered photon is measured by one of the second layer detectors located at forward angles.
This type of polarimeters are suitable for measurements of celestial HG photons because the energy shifts at forward angles are relatively large.

%%% Summary
It is expected to generate HG wave $\gamma$-rays using high-order harmonic radiations from planar undulators and inverse Compton scattering with HG mode laser in the near future.
Because the HG photon could bring large orbital angular momentum like the LG photon, it is as useful as the LG photon to study new interactions with materials.
The present results indicate that one can identify HG wave photons and their node numbers of $n_x$ and $n_y$ by measuring the azimuthal angle dependence of the scattered cross sections for a fixed $\Delta$E or by measuring the energy spectrum of $\Delta$E as a function of the azimuthal angle.

%\noindent
%{\bf Acknowledgements}\\
This work was supported by Grants-in-Aid for Scientific Research of JSPS (JP18H03715, JP16K05360, JP17K05459) and the grant of Joint Research by the National Institutes of Natural Sciences (NINS), (NINS program No, 01111701).

\end{document}